\documentclass[12pt,a4paper]{article}
\usepackage{epsfig}
\begin{document}
\title{Pair production of neutral Higgs bosons from the left-right twin Higgs model at the ILC and LHC}
\author{Wei Ma,  Chong-Xing Yue, Yong-Zhi Wang \\
{\small Department of Physics, Liaoning  Normal University, Dalian,
116029 P. R. China}
\thanks{cxyue@lnnu.edu.cn}}
\date{\today}
\maketitle

\begin{abstract}
\hspace{2mm}In the framework of the left-right twin Higgs model, we
study pair production of the neutral Higgs bosons at the
International Linear Collider ($ILC$) and the $CERN$ $LHC$. We find
that the production cross section of the process
$e^{+}e^{-}\rightarrow \phi^{0}h$ are at the level of several tens
$fb$ at the $ILC$, the production cross section of the
$\phi^{0}\phi^{0}$ pair and $\phi^{0}h$ pair are at the level of
several hundreds $fb$ at the $LHC$. As long as the neutral Higgs
boson $\phi^{0}$ is not too heavy, we conclude that its pair
production might be used to test for the left-right twin Higgs model
at the $LHC$ experiment or in the future $ILC$ experiment.

\vspace{1cm}


\end {abstract}

\vspace{0.8cm}
\newpage

\section*{I. Introduction}
\hspace{0.5cm}The Higgs mechanism is the heart of the standard model
($SM$) providing masses to gauge bosons via electroweak symmetry
breaking ($EWSB$). However, the $SM$ fails to explain the origin of
the fermion mass and has naturalness problems. Many alternative new
physics models with extended Higgs sectors are free from the above
difficulties. The hunt for the Higgs bosons came to be one of the
most important goals for present and future high energy collider
experiments. Apart from the $SM$, neutral Higgs bosons appear in
almost every scenario exploring new phenomena [1]. Pair production
of neutral Higgs bosons at the $CERN$ $LHC$, which will provide a
way to test the Higgs boson self-coupling, may be sensitive to new
physics [2, 3]. Many works have contributed to studies of the
neutral Higgs pair production at the hadron collider in model
independent [4], in $SM$ [5-8], and in new physics models beyond the
$SM$, such as little Higgs models [9], Randall-Sundrum-like models
[10], top condensation models [11], supersymmetric models ($SUSY$)
[12, 13] and models of universal extra dimensions ($UED$) [14].

The $SM$ has been proved by all existing precise experimental data
with its theoretical predictions beyond one-loop level being
coincident with experimental observations. But in the $SM$ the Higgs
boson mass suffers from an instability under radiative corrections,
which is called "hierarchy problem" [15]. Recently, the twin Higgs
mechanism has been proposed as a solution to the little hierarchy
problem. The Higgs bosons emerge as pseudo-Goldstone bosons once the
global symmetry is spontaneously broken. Gauge and $Yukawa$
interactions that break the global symmetry give masses to the
Higgses. The twin Higgs mechanism can be implemented in left-right
models with the additional discrete symmetry being identified with
left-right symmetry [16, 17]. The left-right twin Higgs ($LRTH$)
model contains the $U(4)_{1}\times U(4)_{2}$ global symmetry as well
as the gauged symmetry $SU(2)_{L}\times SU(2)_{R} \times
U(1)_{B-L}$. After Higgs obtained vacuum expectation values ($f$,
$\hat{f}$), the global symmetry $U(4)_{1}\times U(4)_{2}$ breaks
down to $U(3)_{1}\times U(3)_{2}$, and the gauge group
$SU(2)_{R}\times U(1)_{B-L}$ breaks down to the $SM$ $U(1)_{Y}$.
Thus, the $LRTH$ model predicts the existence of the new particles,
such as heavy gauge bosons, heavy scalars, and the top partner $T$,
which can generate rich phenomenology at present and in future
collider experiments [17-21].

In the context of the $LRTH$ model, pair production of the charged
Higgs bosons ($\phi^{+}$, $\phi^{-}$) in the $LRTH$ model at the
$ILC$ and $LHC$ are studied in Ref.[21], but they did not consider
production of the neutral Higgs bosons ($\phi^{0}$, $h$). As we
know, so far production of the neutral Higgs pair at the $LHC$ and
the $ILC$ in the $LRTH$ model has not been considered, which is the
main aim of this paper.

Besides the $SM$-like Higgs boson $h$, there are two additional
neutral Higgs bosons in the $LRTH$ model, which are
$\hat{h}^{0}_{2}$ and $\phi^{0}$. The neutral Higgs boson
$\hat{h}^{0}_{2}$ is a possible dark matter candidate that only
couples to the gauge bosons (including the $SM$ gauge bosons
$\gamma$, $Z$, $W$, and the new gauge boson $Z_{H}$). The production
cross section of $\hat{h}^{0}_{2}$ at the collider is very small and
escapes the detector. Therefore, in this paper, we will not discuss
the production of $\hat{h}^{0}_{2}$ at the $ILC$ or $LHC$. The
neutral Higgs boson $\phi^{0}$ is a pseudoscalar that couples to
both the $SM$ fermions and gauge bosons. The neutral Higgs boson
pair $\phi^{0}h$ can be produced via the processes
$e^{+}e^{-}\rightarrow Z(Z_{H})\rightarrow \phi^{0}h$ at the $ILC$,
and via the partonic processes $q\bar{q}\rightarrow \phi^{0}h
(q=u,c,d,s,b)$, $gg\rightarrow \phi^{0}h$ at the $LHC$,
respectively. While the neutral Higgs pair $\phi^{0}\phi^{0}$ can
only be produced via the partonic process $gg\rightarrow
\phi^{0}\phi^{0}$ and the t-channel partonic process
$b\bar{b}\rightarrow \phi^{0}\phi^{0}$ at the $LHC$. We calculate
all above these processes. Our numerical results denote that, for
$m_{h}=120GeV$, $120GeV\leq m_{\phi^{0}}\leq 180GeV$ and $500GeV
\leq f\leq 1500GeV$: (i)the production cross section of $\phi^{0}h$
at the $ILC$ with the center-of-mass ($c.m.$) energy
$\sqrt{s}=500GeV$ is in the range of $0.92fb \sim 20 fb$; (ii)the
production cross section of $\phi^{0}h$ at the $LHC$ with the $c.m.$
energy $\sqrt{s}=14TeV$ is in the range of $34fb - 306 fb$, and the
main contribution comes from light quarks; (iii)the production cross
section of $\phi^{0}\phi^{0}$ at the $LHC$ is in the range of $4fb -
122fb$, and the main contribution comes from the top quark loop.

This paper is organized as follows. In Sec. II, we briefly review
the essential features of the $LRTH$ model. The relevant couplings
of the neutral Higgs bosons to other particles and the feature of
the decay for the neutral Higgs bosons $\phi^{0}$ are also discussed
in this section. In Secs. III and IV, we give our numerical results
for pair production of neutral Higgs bosons predicted by the $LRTH$
model at the $ILC$ and $LHC$, respectively. Our conclusions are
given in Sec. V.

\section*{II. The LRTH Model}

\vspace*{-0.3cm} \hspace{0.5cm}The $LRTH$ model was first proposed
in Ref.[16] and the details of the model as well as the particle
spectrum, $Feynman$ rules, and some phenomenology analysis have been
studied in Ref.[17]. Here we will briefly review the essential
features of the model and focus our attention on the neutral Higgs
bosons.

The $LRTH$ model is based on the global $U(4)_{1} \times U(4)_{2}$
symmetry with a locally gauged subgroup $SU(2)_{L} \times SU(2)_{R}
\times U(1)_{B-L}$. Two Higgs fields, $H = (H_{L}, H_{R})$ and
$\hat{H} = (\hat{H}_{L}, \hat{H}_{R})$, are introduced and each
transforms as $(4, 1)$ and $(1, 4)$, respectively, under the global
symmetry. $H_{L, R}$ ($\hat{H}_{L, R}$) are two component objects
which are charged under $SU(2)_{L}$ and $SU(2)_{R}$, respectively.
For the gauge couplings $g_{2L}$ and $g_{2R}$ of $SU(2)_{L}$ and
$SU(2)_{R}$, the left-right symmetry implies that $g_{2L}$ =
$g_{2R}$ = $g_{2}$.

The $U(4)_{1}$ [$U(4)_{2}$] group is spontaneously broken down to
its subgroup $U(3)_{1}$ [$U(3)_{2}$] with nonzero vacuum expectation
value ($VEV$) $< H >$ = ($0$, $0$, $0$, $f$) [$< \hat{H}
>$ = ($0$, $0$, $0$, $\hat{f}$)]. The Higgs $VEVs$ also break
$SU(2)_{R} \times U(1)_{B-L}$ down to the $SM$ $U(1)_{Y}$. After
spontaneous global symmetry breaking by $f$ and $\hat{f}$, three
$Goldstone$ bosons are eaten by the new gauge bosons $W_{H}^{\pm}$
and $Z_{H}$. After the $SM$ electroweak symmetry breaking, the three
additional $Goldstone$ bosons are eaten by the $SM$ gauge bosons
$W^{\pm}$ and $Z$.

The fermion sector of the $LRTH$ model is similar to that of the
$SM$, with the right-handed quarks ($u_{R}$, $d_{R}$) and leptons
($l_{R}$, $\upsilon_{R}$) form fundamental representations of
$SU(2)_{R}$. In order to give the top-quark mass of the order of the
electroweak scale, a pair of vectorlike quarks $Q_{L}$ and $Q_{R}$
are introduced. The mass eigenstates, which contain one the $SM$ top
quark $t$ and a heavy top partner $T$, are mixtures of the gauge
eigenstates. Their masses are given by \vspace{-0.25cm}
\begin{eqnarray}
m_{t}^{2} = \frac{1}{2}(M^{2}+y^{2}f^{2}-N_{t}),\hspace{0.5cm}
M_{T}^{2} = \frac{1}{2}(M^{2}+y^{2}f^{2}+N_{t}).
\end{eqnarray}
where $N_{t} = \sqrt{(y^{2}f^{2}+M^{2})^{2}-y^{4}f^{4}\sin^{2}2x}$
with $x=\nu/\sqrt{2}f$, in which $\nu=246GeV$ is the scale of the
$EWSB$. Provided $M_{T} \leq f$ and that the parameter $y$ is of
order one, the top $Yukawa$ coupling will also be of order one. The
parameter $M$ is essential to the mixing between the $SM$ top quark
and its partner $T$.

According the symmetry-breaking pattern discussed above, with
certain reparametrizations of the fields, there are left with four
Higgs bosons in the $LRTH$ spectrum that couple to both the fermion
sector and the gauge boson sector. They are one neutral Higgs bosons
$\phi^{0}$, a pair of charged Higgs bosons $\phi^{\pm}$, and the
$SM$-like physical Higgs $h$. In addition, there is an $SU(2)_{L}$
doublet $\hat{h} = (\hat{h}_{1}^{+}, h_{2}^{0})$ that couples to the
gauge boson sector only (including the $SM$ gauge bosons $\gamma$,
$Z$, $W$, and the new gauge boson $Z_{H}$). The lightest particle in
$\hat{h}$, typically one of the neutral components, is stable, and
therefore constitutes a good dark matter candidate.

These neutral Higgs bosons can couple to each others, and also can
couple to the ordinary fermions, ordinary gauge bosons, new top
quark T, and new gauge boson $Z_{H}$. The couplings expression forms
which are related our calculation, are shown as [17]

\vspace{-0.8cm}
\begin{eqnarray}
\nonumber
&&\hspace*{0.03cm}\phi^{0}\bar{d}_{1,2,3}d_{1,2,3}:im_{d_{i}}\gamma_{5}/(\sqrt{2}f);
\hspace*{2cm}\phi^{0}\bar{u}_{1,2}u_{1,2}\hspace*{-0.1cm}:\hspace*{-0.1cm}-im_{u_{i}}\gamma_{5}/(\sqrt{2}f)\hspace*{0.05cm};\\
\nonumber
&&ht\bar{t}\hspace*{0.3cm}:-\hspace*{0.05cm}em_{t}C_{L}C_{R}\hspace*{0.05cm}/\hspace*{0.1cm}(2m_{W}S_{W})
\hspace*{0.05cm};\hspace*{2.05cm}
\phi^{0}\bar{t}t\hspace*{0.31cm}:\hspace*{-0.1cm}-iyS_{R}S_{L}\gamma_{5}/\sqrt{2}\hspace*{0.08cm};
\\ \nonumber
&&hT\bar{T}:-y(S_{R}S_{L}-C_{L}C_{R}x)/\sqrt{2};\hspace*{2cm}
\phi^{0}\bar{T}T:\hspace*{-0.1cm}-iyC_{L}C_{R}\gamma_{5}/\sqrt{2};
\\ \nonumber
&&h\phi^{0}\phi^{0}\hspace*{-0.15cm}:\hspace*{-0.1cm}x(\hspace*{-0.05cm}30p_{2}\hspace*{-0.1cm}\cdot\hspace*{-0.1cm}
p_{3}\hspace*{-0.1cm}+\hspace*{-0.1cm}11p_{1}\hspace*{-0.1cm}\cdot\hspace*{-0.1cm}
p_{1}\hspace*{-0.1cm})\hspace*{-0.05cm}/\hspace*{-0.05cm}(\hspace*{-0.05cm}27\sqrt{2}f\hspace*{-0.05cm});
\hspace*{1.6cm}h\phi^{0}Z_{\mu}\hspace*{-0.1cm}:\hspace*{-0.05cm}i
e x p_{3\mu}\hspace*{-0.05cm}/\hspace*{-0.05cm}(\hspace*{-0.05cm}6
C_{W} S_{W})\hspace*{-0.05cm};
\\ \nonumber
&&Z_{H\mu}\bar{u}_{1,2}u_{1,2}\hspace*{-0.1cm}:\hspace*{-0.1cm}- e
\gamma_{\mu}(2S_{W}^{2}P_{L}\hspace*{-0.08cm}+\hspace*{-0.08cm}(1-7cos2\theta_{W})P_{R})/(12C_{W}S_{W}
\sqrt{cos2\theta_{W}});
\\ \nonumber
&&Z_{H\mu}\bar{d}_{1,2,3}d_{1,2,3}:\hspace*{0.1cm}-\hspace*{0.1cm}
e\hspace*{0.05cm}
\gamma_{\mu}\hspace*{0.05cm}(S_{W}^{2}P_{L}+(3-5S_{W}^{2})P_{R})/(6C_{W}S_{W}\sqrt{cos2\theta_{W}})\hspace*{0.03cm};
\\
&&h\phi^{0}Z_{H\mu}:i e x((14-17S_{W}^{2})p_{2\mu}-(4-
S_{W}^{2})p_{1\mu})/(18 S_{W}C_{W}\sqrt{cos2\theta_{W}}).
\end{eqnarray}
Where $p_{1}$, $p_{2}$, and $p_{3}$ refer to the incoming momentum
of the first, second and third particles, respectively. $u_{i}$ and
$d_{i}$ represent the upper- and down- type fermions, respectively.
$S_{W}= \sin\theta_{W}$, $C_{W}=cos\theta_{W}$, and $\theta_{W}$ is
the $Weinberg$ angle. At the leading order of $1/f$, the sine values
of the mixing angles $\alpha_{L}$ and $\alpha_{R}$ can be written as

\vspace*{-0.7cm}
\begin{eqnarray}
\hspace*{10mm}S_{L}=\sin\alpha_{L}\simeq\frac{M}{M_{T}}\sin
x,\hspace{0.8cm}
S_{R}=\sin\alpha_{R}\simeq\frac{M}{M_{T}}(1+\sin^{2}x).
\end{eqnarray}
$C_{L}$ and $C_{R}$ are the cosine values of the mixing angles
$\alpha_{L}$ and $\alpha_{R}$, respectively. $P_{L(R)} =
(1\mp\gamma_{5})/2$ is the left (right)-handed projection operator.

In the framework of the $LRTH$ model, the mass of the neutral Higgs
boson $\phi^{0}$ can be anything below $f$ here we consider another
possibility, in which the mass is around 150GeV [17]. Similar to the
$SM$ Higgs boson, $\phi^{0}$ can decay to $\gamma\gamma$ through the
top-quark loop and heavy top-quark loop. But unlike the $SM$ Higgs
boson, in the $LRTH$ model, the light neutral Higgs boson $\phi^{0}$
is a pseudoscalar boson, due to its pseudoscalar nature, there is no
$\phi^{0}WW$ and $\phi^{0}ZZ$ couplings at tree level. So, the
one-loop $SM$ gauge boson contribution to $\phi^{0}\gamma\gamma$ is
zero. In general, the light neutral Higgs boson $\phi^{0}$ decays
into $b\bar{b}$, $c\bar{c}$, $\tau^{+}\tau^{-}$, $gg$ and
$\gamma\gamma$. Now we discuss the branching ratios for the possible
decay modes of $\phi^{0}$. The decay width of $\phi^{0}\rightarrow
f\bar{f}$ is proportional to the square of the corresponding
$Yukawa$ coupling, with an additional suppression factor of
$\nu^{2}/(2f^{2})$ comparing to that of the $SM$ Higgs boson. The
concrete expressions of the decay widths for the different decay
channels are given as follows:

\vspace{-0.8cm}
\begin{eqnarray}
\nonumber&&\Gamma(\phi^{0}\rightarrow
b\bar{b})\hspace{0.2cm}=\hspace{0.18cm}\frac{3G_{F}m_{\phi^{0}}\nu^{2}m_{b}^{2}}{8\sqrt{2}\pi
f^{2}}(1-4m_{b}^{2}/m_{\phi^{0}}^{2})^{\frac{3}{2}},
\\ \nonumber
&&\Gamma(\phi^{0}\rightarrow
c\bar{c})\hspace{0.2cm}=\hspace{0.18cm}\frac{3G_{F}m_{\phi^{0}}\nu^{2}m_{c}^{2}}{8\sqrt{2}\pi
f^{2}},
\\ \nonumber
&&\Gamma(\phi^{0}\rightarrow
\tau^{+}\tau^{-})=\frac{G_{F}m_{\phi^{0}}\nu^{2}m_{\tau}^{2}}{8\sqrt{2}\pi
f^{2}},
\\ \nonumber
&&\Gamma(\phi^{0}\rightarrow\gamma\gamma)=\frac{G_{F}\alpha^{2}m_{\phi^{0}}^{3}}{128\sqrt{2}\pi^{3}}|
\sum_{f}N_{c}^{f}Q_{f}^{2}A_{f}^{\phi^{0}}(\tau_{f})|^{2},
\\
&&\Gamma(\phi^{0}\rightarrow
gg)\hspace{0.08cm}=\hspace{0.14cm}\frac{G_{F}^{2}\alpha_{s}^{2}m_{\phi^{0}}^{3}}
{48\sqrt{2}\pi^{3}}|\sum_{q}A_{q}^{\phi^{0}}(\tau_{q})|^{2}.
\end{eqnarray}
Where $m_{b}$, $m_{c}$, and $m_{\tau}$ are the masses of the $SM$
fermions $b$, $c$ and $\tau$, respectively. The index $f$
corresponds to $q$ and $l$ ($q=quark$, $l=lepton$). Where
$N_{c}^{f}=1,3$ for $f=l,q$, respectively. $Q_{f}$ is the charge of
the fermion $f$. Similar with Ref.[22], the function
$A_{f}^{\phi^{0}}$ can be written as:

\vspace{-1cm}
\begin{eqnarray}
A_{f}^{\phi^{0}}=2\tau_{f}[1+(1-\tau_{f})f(\tau_{f})].
\end{eqnarray}
where $\tau_{f}=4m_{f}^{2}/m_{\phi^{0}}^{2}$. In Ref.[22], the
function $f(\tau_{f})$ has two parts corresponding to the
$\tau_{f}\geq1$ and $\tau_{f}<1$ two conditions. In our numerical
estimation, we have neglected the contributions of the light
fermions. Therefore, in the $LRTH$ model, there is
$\tau_{t(T)}=4m_{t(T)}^{2}/m_{\phi^{0}}^{2}\geq 1$, and the function
$f(\tau_{f})$ is given by \vspace{-0.15cm}
\begin{eqnarray}
f(\tau)=arcsin^{2}\frac{1}{\sqrt{\tau}}.
\end{eqnarray}

\vspace*{-0.45cm}\hspace*{-0.75cm} where $A_{q}^{\phi^{0}}$,
$\tau_{q}$, and $f(\tau_{q})$ in Eq.(4) are defined the same as
$A_{f}^{\phi^{0}}$, $\tau_{f}$ and $f(\tau_{f})$, but only for
quarks.

Using above partial widths of the neutral Higgs boson $\phi^{0}$,
its total width $\Gamma$ can be approximately written as

\vspace{-1.2cm}
\begin{eqnarray}
\hspace{0.8cm}\Gamma=\Gamma_{b\bar{b}}+\Gamma_{c\bar{c}}+\Gamma_{\tau^{+}\tau^{-}}+
\Gamma_{\gamma\gamma}+\Gamma_{gg} .
\end{eqnarray}

\begin{figure}[htb]
\begin{center}
\vspace{-1.4cm}\hspace*{-1cm}\epsfig{file=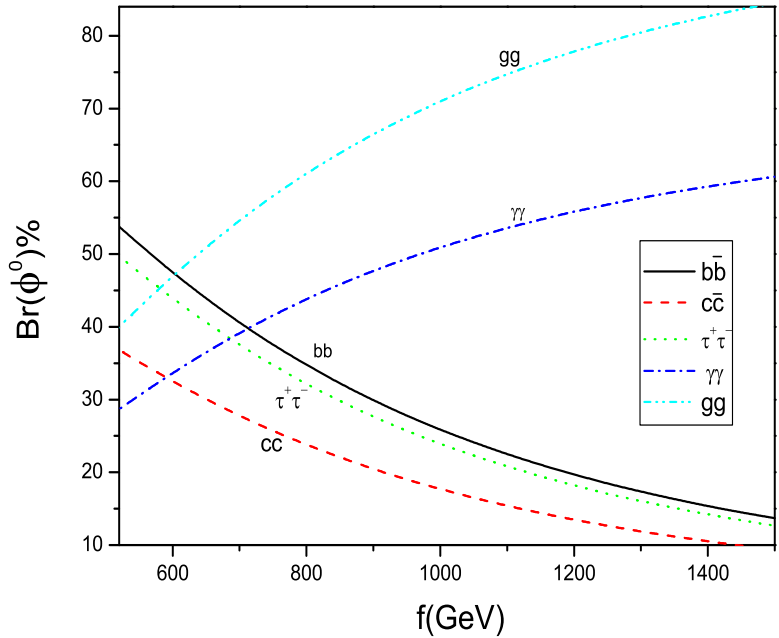,width=345pt,height=300pt}
\vspace{-1.2cm} \hspace*{-10mm}\caption{The branching ratios of the
neutral Higgs boson $\phi^{0}$ for different decay modes
\hspace*{1.8cm}as functions of the free parameter $f$ for
$M=150GeV$, $m_{\phi^{0}}=120GeV$. In \hspace*{1.8cm}order to see
the trend clearly, we have multiplied $Br(\phi^{0}\rightarrow
c\bar{c}$), $Br(\phi^{0}\rightarrow\tau^{+}\tau^{-}$)
\hspace*{1.8cm}, and $Br(\phi^{0}\rightarrow\gamma\gamma$) by the
factors 10, 20, and 300, respectively.} \label{ee}
\end{center}
\vspace{-0.55cm}
\end{figure}

We summed up our numerical results of the branching ratios of the
neutral Higgs boson $\phi^{0}$ for different decay modes
$Br(\phi^{0})$ in $Fig.1$. To get the numerical results, the $SM$
parameters involved are taken as $m_{b}=4.8GeV$, $m_{c}=1.25GeV$ and
$m_{\tau}=1.78GeV$ [23]. In $Fig.1$, we plot $Br(\phi^{0})$ as a
function of free parameter $f$ for $M=150GeV$ and
$m_{\phi^{0}}=120GeV$. One can see from $Fig.1$ that the decay
branching ratios of $\phi^{0}$ are sensitive to the parameter $f$.
If we assume that the parameter $f$ is in the range of $500GeV\sim
1500GeV$, the value of the branching ratio $Br(\phi^{0}\rightarrow
b\bar{b})$ is in the range of $14\%- 55\%$, and the branching ratio
$Br(\phi^{0}\rightarrow gg)$ is in the range of $38\%- 85\%$. The
values of $Br(\phi^{0}\rightarrow c\bar{c}$),
$Br(\phi^{0}\rightarrow \tau^{+}\tau^{-}$) and
$Br(\phi^{0}\rightarrow\gamma\gamma$) are much smaller than those of
$Br(\phi^{0}\rightarrow b\bar{b})$ and $Br(\phi^{0}\rightarrow gg$).
Therefore, in order to see the trend clearly, in $Fig.1$ we have
multiplied them by 10, 20 and 300, respectively. The real numerical
results are $Br(\phi^{0}\rightarrow c\bar{c}$)$=0.9\%- 3.8\%$,
$Br(\phi^{0}\rightarrow \tau^{+}\tau^{-}$)$=0.6\%- 2.6\%$, and
$Br(\phi^{0}\rightarrow \gamma\gamma$)$=0.09\%- 0.2\%$. Our
numerical results agree quite well with Ref.[17], in that the
branching ratio $Br(\phi^{0}\rightarrow\gamma\gamma)$ is roughly
same as $Br(h\rightarrow\gamma\gamma$) for $m_{h}=m_{\phi^{0}}$.

\vspace{-0.2cm}
\section*{III. \hspace*{-0.05cm}Pair\hspace*{-0.1cm} production \hspace*{-0.1cm}of \hspace*{-0.2cm}
neutral\hspace*{-0.1cm} Higgs \hspace*{-0.1cm}bosons
\hspace*{-0.1cm}at\hspace*{-0.1cm} the\hspace*{-0.1cm} ILC}

\vspace{-0.2cm} \hspace{0.5cm}In many cases, the $ILC$ can
significantly improve the $LHC$ measurements. If a Higgs boson is
discovered, it will be crucial to determine its couplings with high
accuracy, to understand the so-called mechanism of $EWSB$ [24]. The
high resolution profile determination of a light Higgs boson (mass,
couplings, self-couplings, etc.) can be carried out at the $ILC$,
where clear signals of Higgs events are expected with backgrounds
that can be reduced to a magnitude level. With the $LHC$ guidance,
the $ILC$, which is currently being designed, will further improve
our knowledge of the Higgs sector if that is how nature decided to
create mass [24]. It was demonstrated in Ref.[25] that physics at
the $LHC$ and at the $ILC$ will be complementary to each other in
many respects. So far, many works have been contributed to studies
of the neutral Higgs boson pair production at the $ILC$, in the $SM$
[26-28] and in new physics beyond the $SM$ [29-32].

From the discussions given in Sec. II, we can see that the neutral
Higgs boson pair $\phi^{0}\phi^{0}$ cannot be produced exclusively
at the $ILC$ because $\phi^{0}\phi^{0}$can not couple with gauge
boson $Z$ or $Z_{H}$. However, the neutral Higgs boson pair
$\phi^{0}h$ can be produced via the processes $e^{+}e^{-}\rightarrow
Z(Z_{H}) \rightarrow \phi^{0}h$ at the $ILC$. The $Feynman$ diagrams
of the process $e^{+}(p_{1})e^{-}(p_{2})\rightarrow
\phi^{0}(p_{3})h(p_{4})$ are shown in $Fig$.2.

\begin{figure}[htb]
\vspace{-6.3cm}
\begin{center}
\hspace*{-0.5cm}\epsfig{file=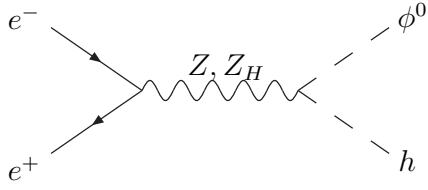,width=580pt,height=800pt}
\vspace{-20.7cm}\caption{Feynman diagrams for the process
$e^{+}e^{-}\rightarrow \phi^{0}h$.}
\end{center}
\vspace{-0.7cm}
\end{figure}

At the leading order, the production amplitude of the process can be
written as
\begin{eqnarray}
M_{1}=M_{Z}+M_{Z_{H}}
\end{eqnarray}
\vspace*{-1.1cm}
 with
\begin{eqnarray} \nonumber
\hspace*{2.3cm}M_{Z}\hspace*{-0.55cm}&&=\frac{e^{2}x(-1+4S_{W}^{2})}{24C_{W}^{2}S_{W}^{2}}\bar{v}_{e}(p_{2})\frac{p_{12}\hspace*{-0.5cm}/}
{p_{12}^{2}-m_{Z}^{2}}u_{e}(p_{1})
\\ \nonumber
\hspace*{2.3cm}&&+\frac{e^{2}x}{24C_{W}^{2}S_{W}^{2}}\bar{v}_{e}(p_{2})\frac{p_{12}\hspace*{-0.5cm}/}{p_{12}^{2}-m_{Z}^{2}}
\gamma_{5}u_{e}(p_{1}),
\\ \nonumber
\hspace*{2.3cm}M_{Z_{H}}\hspace*{-0.7cm}&&=\frac{-e^{2}x(14-17S_{W}^{2})}{36C_{W}^{2}cos2\theta_{W}}
\bar{v}_{e}(p_{2})\frac{p_{3}\hspace*{-0.35cm}/}{p_{12}^{2}-m_{Z_{H}}^{2}}P_{L}u_{e}(p_{1})
\\ \nonumber
\hspace*{2.3cm}&&+\frac{-e^{2}x(14-17S_{W}^{2})(1-3C_{W}^{2})}{72C_{W}^{2}S_{W}^{2}cos2\theta_{W}}
\bar{v}_{e}(p_{2})\frac{p_{3}\hspace*{-0.35cm}/}{p_{12}^{2}-m_{Z_{H}}^{2}}P_{R}u_{e}(p_{1})
\\ \nonumber
\hspace*{2.3cm}&&+\frac{e^{2}x(4-S_{W}^{2})}{36C_{W}^{2}cos2\theta_{W}}
\bar{v}_{e}(p_{2})\frac{p_{4}\hspace*{-0.35cm}/}{p_{12}^{2}-m_{Z_{H}}^{2}}P_{L}u_{e}(p_{1})
\\ \nonumber
\hspace*{2.3cm}&&+\frac{e^{2}x(1-3C_{W}^{2})(4-S_{W}^{2})}{72C_{W}^{2}S_{W}^{2}cos2\theta_{W}}
\bar{v}_{e}(p_{2})\frac{p_{4}\hspace*{-0.35cm}/}{p_{12}^{2}-m_{Z_{H}}^{2}}P_{R}u_{e}(p_{1}).\\
\nonumber
\end{eqnarray}

\vspace*{-0.7cm}\hspace*{-0.7cm} where $p_{12}$ is the momentum of
the propagator, which is the sum of the incoming momentums $p_{1}$
and $p_{2}$. With the above production amplitudes, we can obtain the
production cross section directly.

From the above discussions, we can see that, except for the $SM$
input parameters $\alpha=1/128.8$, $S_{W}=\sqrt{0.2315}$,
$m_{Z}=91.1876GeV$, $m_{h}=120GeV$ [23], the cross section $\sigma$
of pair production for the neutral Higgs boson $\phi^{0}h$ at the
$ILC$ is dependent on the model dependent parameters $f$ and
$m_{\phi^{0}}$. In our numerical estimation, we will assume that the
values of the free parameters $f$ and $m_{\phi^{0}}$ are in the
ranges of $500GeV- 1500GeV$ and $100GeV- 180GeV$, respectively.

In $Fig.$3, we plot the production cross section $\sigma$ of the
process $e^{+}e^{-}\rightarrow \phi^{0}h$ as a function of the scale
parameter $f$ for the $c.m.$ energy $\sqrt{s}=500GeV$,
$m_{h}=120GeV$ and three values of $m_{\phi^{0}}$. We can see that
$\sigma$ is sensitive to the scale parameter $f$ and the mass
parameter $m_{\phi^{0}}$. For $500GeV \leq f \leq 1500GeV$ and
$120GeV \leq m_{\phi^{0}} \leq180GeV$, its value is in the range of
$0.92fb- 20fb$. According to an update of parameter for $ILC$ at
2006 [33]\footnote{Thanks to the referees for offering this
reference to us.}, one can see that, an integrated luminosity of
$500fb^{-1}$ should be achieved in the first four years of running
after one year of commissioning. Therefore, if we assume the
integrated luminosity for the $ILC$ is $500fb^{-1}$, there will be
$10^{2}- 10^{4}$ $\phi^{0}h$ events to be generated at the $ILC$.

\begin{figure}[htb]
\begin{center}
\vspace{-1.2cm}\epsfig{file=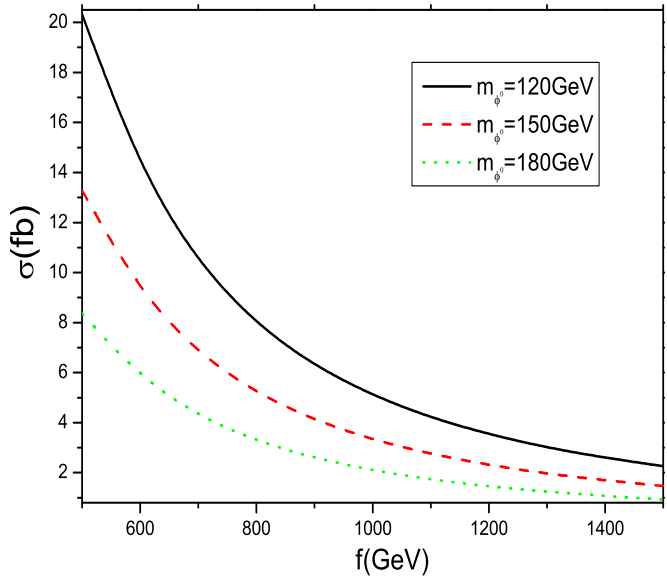,width=295pt,height=270pt}
\vspace{-1cm} \hspace*{-3mm}\caption{The production cross section
$\sigma$ of $e^{+}e^{-}\rightarrow \phi^{0}h$ as a function of the
parameter \hspace*{18mm}$f$ for three values of $m_{\phi^{0}}$,
$m_{h}=120GeV$, and the $c.m.$ energy $\sqrt{s}=500GeV$.} \label{ee}
\end{center}
\vspace{-0.7cm}
\end{figure}

From the discussions given in Sec. II, we can see that the possible
decay modes of the neutral Higgs boson $\phi^{0}$ are $b\bar{b}$,
$c\bar{c}$, $\tau^{+}\tau^{-}$, $gg$ and $\gamma\gamma$. The
$SM$-like neutral Higgs boson $h$ has similar decay features with
those of $\phi^{0}$. Therefore, the signatures of neutral Higgs
boson pair $\phi^{0}h$ is similar to those of the neutral Higgs
boson pair $\phi^{0}\phi^{0}$ at the high energy colliders.  From
the numerical results given in Sec. II, one can see that, for the
masses $m_{\phi^{0}}\leq 180GeV$, the possible signals of
$\phi^{0}h$ can be seen as four $b$ quarks,
\begin{eqnarray}
e^{+}e^{-}\rightarrow \phi^{0}h\rightarrow b\bar{b}b\bar{b}.
\end{eqnarray}

The production rate of the $b\bar{b}b\bar{b}$ final state in the
$LRTH$ model can be easily estimated using the formula
$\sigma_{s}=\sigma \times Br(\phi^{0}\rightarrow b\bar{b})\times
Br(h\rightarrow b\bar{b})$. If we assume the integrated luminosity
$\pounds_{int}=500fb^{-1}$ for the $ILC$ with the $c.m.$ energy
$\sqrt{s}=500GeV$, then there will be $9- 3.0\times10^{3}$
$b\bar{b}b\bar{b}$ events to be generated at the $ILC$, which is
significantly larger than that for  the $SM$ Higgs boson pair
production process $e^{+}e^{-}\rightarrow hh\rightarrow
b\bar{b}b\bar{b}$ [26-28].
 Therefore, we hope that by using very
efficient $\mu$-vertex detectors to tag the $b$ quark jets, we might
detect the possible signatures of the neutral Higgs boson $\phi^{0}$
via the process $e^{+}e^{-}\rightarrow \phi^{0}h$ in the future
$ILC$ experiments. Certainly, detailed confirmation of the
observability of the signals generated by the process
$e^{+}e^{-}\rightarrow Z(Z_{H})\rightarrow \phi^{0}h$ would require
Monte-Carlo simulations of the signals and backgrounds, which is
beyond the scope of this paper.

\vspace{-0.5cm}
\section*{IV. \hspace*{-0.05cm}Pair\hspace*{-0.1cm} production \hspace*{-0.1cm}of \hspace*{-0.15cm}
neutral\hspace*{-0.1cm} Higgs \hspace*{-0.1cm}bosons
\hspace*{-0.1cm}at\hspace*{-0.1cm} the\hspace*{-0.1cm} LHC}

\hspace{0.5cm}The $LHC$ has a good potential for discovery of a
neutral Higgs boson. Now we look at pair production of the neutral
Higgs bosons predicted by the $LRTH$ model at the $LHC$. From the
above discussions, we can see that both the $\phi^{0}\phi^{0}$ pair
and $\phi^{0}h$ pair can be produced at the $LHC$. In this section,
we will consider both of these cases.

\vspace{0.3cm} \hspace{-0.6cm}\textbf{A. $\phi^{0}\phi^{0}$ pair
production}

\vspace{0.1cm}First, we study production of the neutral Higgs boson
pair $ \phi^{0}\phi^{0}$ at the $LHC$. At the $LHC$, the neutral
Higgs boson pair $\phi^{0}\phi^{0}$ can be produced through two
mechanisms. One is loop-induced production via gluon fusion
($gg\rightarrow \phi^{0}\phi^{0}$) and the other is from the
t-channel quark-antiquark annihilation ($q\bar{q}\rightarrow
\phi^{0}\phi^{0}$). The relevant $Feynman$ diagrams are shown in
$Fig.4$. Considering the couplings of the neutral Higgs boson $
\phi^{0}$ to the $SM$ fermions are proportional to the factor of
$m_{q}/f$ and the smallness masses of the quarks $q=u, c, d,$ and
$s$, we have neglected their contributions to production of the
neutral Higgs boson pair $\phi^{0}\phi^{0}$.

In this paper, we calculate all production channels for the neutral
Higgs boson pair $\phi^{0}\phi^{0}$ at the $LHC$, as shown in
$Fig.$4, including triangle diagrams, box diagrams and tree-level
diagram. Each loop diagram is composed of some scalar loop
functions, which are calculated by using $LoopTools$ [34]. The
hadronic cross section at the $LHC$ is obtained by convoluting the
partonic cross sections with the parton distribution functions
($PDFs$). In our numerical calculation, we will use CTEQ6L $PDFs$
for the gluon and quark $PDFs$ [35]. The renormalization scale
$\mu_{R}$ and the factorization scale $\mu_{F}$ are chosen to be
$\mu_{R}=\mu_{F}=2m_{\phi^{0}}$. Because the calculation of the loop
diagrams are too tedious and the analytical expression are lengthy,
we will not present those here.

\begin{figure}[htb]
\vspace{-4.8cm}
\begin{center}
\hspace*{1cm}\epsfig{file=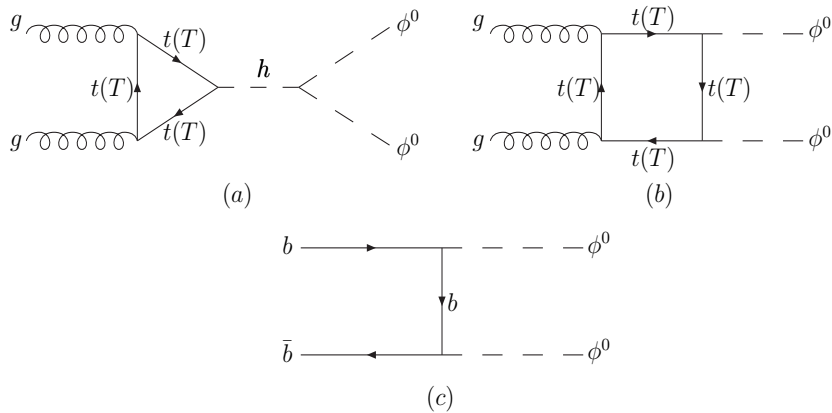,width=450pt,height=680pt}
\vspace{-14.5cm} \caption{ One-loop $Feynman$ diagrams for the
subprocess $gg\rightarrow\phi^{0}\phi^{0}$(a,b) and
tree-\hspace*{20mm}level $Feynman$ diagram for the subprocess
$b\bar{b}\rightarrow \phi^{0}\phi^{0}(c)$ in the $LRTH$
\hspace*{20mm}model. The diagrams obtained by exchanging the two
gluons or exchanging \hspace*{20mm}the two Higgs bosons are not
shown here.}
\end{center}
\vspace{-0.65cm}
\end{figure}

It is obvious that the production cross section $\sigma$ of the
neutral Higgs boson pair $\phi^{0}\phi^{0}$ at the $LHC$ are
dependent on the model dependent parameters $f$, $m_{\phi^{0}}$, and
$M$. Similar to the calculation at the $ILC$, we assume that the
values of the free parameters $f$ and $m_{\phi^{0}}$ are in the
ranges of $500GeV- 1500GeV$ and $100GeV- 180GeV$, respectively.
Besides, we assume the mixing parameter $M$ is in the range of
$100GeV- 200GeV$. Our numerical results are summarized in $Figs.5$
and $6$.

To see contributions of the different partonic processes to the
total hadronic cross section, we plot the total and partial hadronic
cross sections for different partonic processes as functions of the
scale parameter $f$  for the parameters $M=100GeV$ and
$m_{\phi^{0}}=120GeV$ in $Fig.$5. We see from $Fig.$5 that
production of the neutral Higgs boson pair $\phi^{0}\phi^{0}$ is
dominated by the partonic process $gg\rightarrow \phi^{0}\phi^{0}$
induced by the top-quark loop diagrams. For $M=100GeV$,
$m_{\phi^{0}}=120GeV$, and $500GeV \leq f \leq 150GeV $, the value
of the total production cross section is in the range of $4fb\sim
122fb$, and the value of the production cross section coming from
the top-quark loop diagrams is in the range of $1.5fb- 105fb$. This
is because the contributions of the box diagrams are generally much
smaller than those of the triangle diagrams, and furthermore the
coupling $ht\bar{t}$ is much larger than the coupling $hT\bar{T}$ or
the coupling $\phi^{0}b\bar{b}$. If we assume the integrated
luminosity $\pounds_{int}=100fb^{-1}$ for the $LHC$ with the $c.m.$
energy $\sqrt{s}=14TeV$, then there will be $4\times10^{2}-
1.22\times10^{4}$ events to be generated at the $LHC$.

\begin{figure}[htb]
\vspace{-0.9cm}
\begin{center}
\epsfig{file=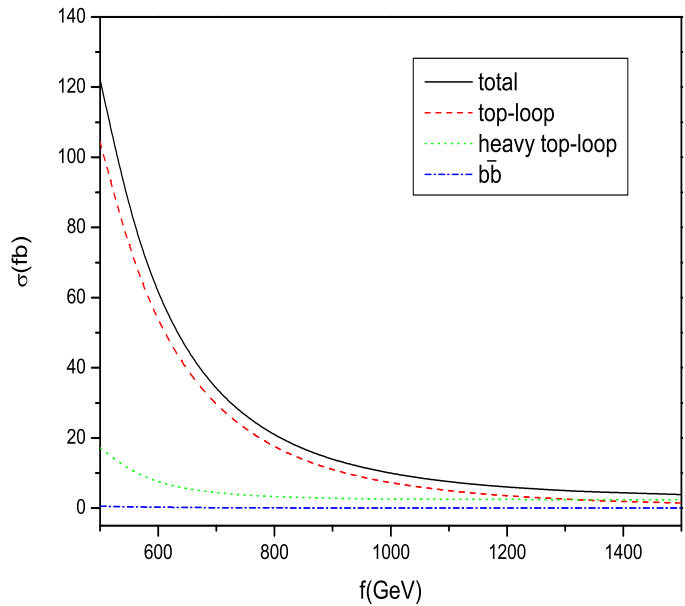,width=300pt,height=280pt}
\vspace{-1.1cm}\hspace{-0.5cm} \caption{The total and partial
hadronic cross sections for different partonic processes
\hspace*{1.9cm}as functions of the free parameter $f$ for the
parameters $M=100GeV$ and \hspace*{1.9cm}$m_{\phi^{0}}=120GeV$.}
\label{ee}
\end{center}
\vspace*{-0.75cm}
\end{figure}

In order to see the effects of the mass parameter $m_{\phi^{0}}$ on
the total cross section $\sigma$, we plot $\sigma$ as a function of
$m_{\phi^{0}}$ for  $f=500GeV$ and three values of the mixing
parameter $M$ in $Fig.$6. One can see from $Fig.$6 that the total
cross section $\sigma$ is sensitive to the mass parameter
$m_{\phi^{0}}$, while is not sensitive to the mixing parameter $M$.
This is because $M$ is introduced to generate the mass mixing term
$Mq_{L}q_{R}$, which is included in the gauge invariant top $Yukawa$
terms allowed by gauge invariance. From the relevant $Feynman$ rules
we can see that, the mixing parameter $M$ does not influence the
production cross section $\sigma$ of the neutral Higgs boson
$\phi^{0}$ too much. For $f=500GeV$, $M=200GeV$, and
$m_{\phi^{0}}=100GeV- 180GeV$, the total cross section $\sigma$ is
in the range of $16fb- 253fb$.

\begin{figure}[htb]
\vspace{-0.8cm}
\begin{center}
\epsfig{file=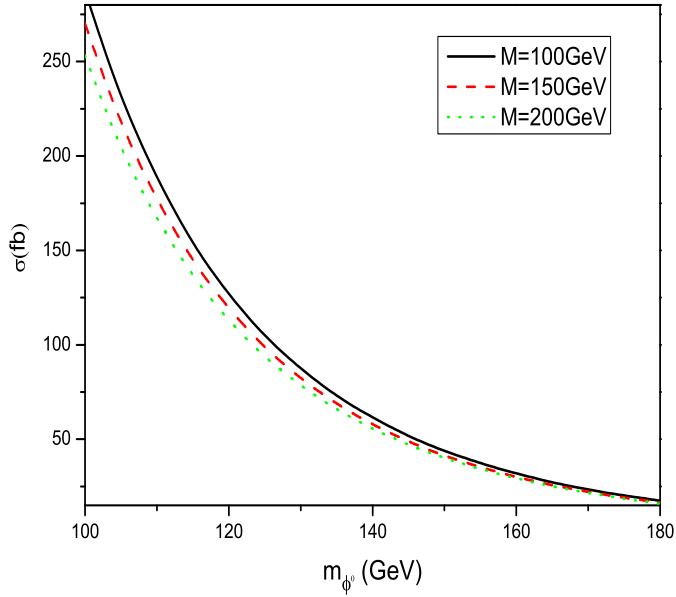,width=292pt,height=280pt}
\vspace{-1.1cm}\hspace{-0.7cm} \caption{The total production cross
section $\sigma$ as a function of free parameter $m_{\phi^{0}}$ for
\hspace*{1.8cm} three values of mixing parameter $M$.} \label{ee}
\end{center}
\vspace{-1.1cm}
\end{figure}

 \vspace{0.4cm}\hspace{-0.6cm}\textbf{B. $\phi^{0}h$ pair
production}

\vspace{0.1cm}Now we consider production of the neutral Higgs boson
pair $ \phi^{0}h$ at the $LHC$. At the $LHC$, the neutral Higgs
boson pair $\phi^{0}h$ can be mainly produced through two
mechanisms: (i) $q\bar{q}\rightarrow \phi^{0}h$, where $q=u, d, c,
s, b$; (ii) the loop-induced gluon fusion process $gg\rightarrow
\phi^{0}h$. The relevant $Feynman$ diagrams are shown in $Fig.$7.

\begin{figure}[htb]
\vspace{-1.8cm}
\begin{center}
\hspace*{1cm}\epsfig{file=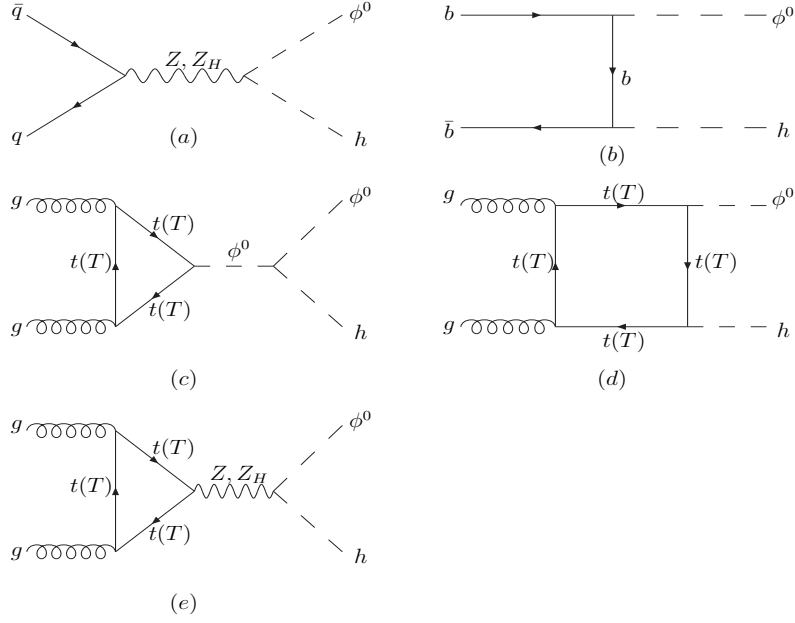,width=440pt,height=550pt}
\vspace{-10.5cm} \caption{Tree-level $Feynman$ diagrams for the
process $q\bar{q}\rightarrow \phi^{0}h (q=u, d, c, s, b)$ (a,b)
\hspace*{1.9cm}and one-loop $Feynman$ diagrams for
$gg\rightarrow\phi^{0}h$ (c,d,e) in the $LRTH$ model.}
\end{center}
\vspace{-0.65cm}
\end{figure}

Using the relevant $Feynman$ rules, we can write the invariant
amplitude for the partonic process
$q(p_{1})\bar{q}(p_{2})\rightarrow \phi^{0}(p_{3})h(p_{4})$ as
\begin{eqnarray}
\nonumber &&M_{2}(q)=M_{21}(q),\hspace*{2.2cm}for\hspace*{0.1cm} q=u,\hspace*{0.1cm}c\\
\nonumber &&M_{2}(q)=M_{22}(q),\hspace*{2.2cm}for\hspace*{0.1cm} q=d,\hspace*{0.1cm}s\\
&&M_{2}(q)=M_{22}(q)+M_{23}(q),\hspace*{0.5cm}for \hspace*{0.1cm}q=b
\end{eqnarray}

For the s-channel partonic processes $q\bar{q}\rightarrow
Z(Z_{H})\rightarrow \phi^{0}h$ ($q=u$ and $c$), the invariant
amplitude can be written
\begin{eqnarray}
\nonumber
M_{21}(q)&=&(\frac{-e^{2}x}{24S_{W}^{2}C_{W}^{2}}+\frac{e^{2}x}{18C_{W}^{2}})
\bar{v}(p_{2})\frac{p_{12}\hspace*{-0.5cm}/}{p_{12}^{2}-m_{Z}^{2}}P_{L}u(p_{1})
\\ \nonumber
&+&\frac{e^{2}x}{18C_{W}^{2}}\bar{v}(p_{2})\frac{p_{12}\hspace*{-0.5cm}/}{p_{12}^{2}-m_{Z}^{2}}P_{R}u(p_{1})
\\ \nonumber
&+&\frac{e^{2}x(14-17S_{W}^{2})}{108C_{W}^{2}cos2\theta_{W}}
\bar{v}(p_{2})\frac{p_{3}\hspace*{-0.35cm}/}{p_{12}^{2}-m_{Z_{H}}^{2}}P_{L}u(p_{1})
\\ \nonumber
&+&\frac{-e^{2}x(4-S_{W}^{2})}{108C_{W}^{2}cos2\theta_{W}}
\bar{v}(p_{2})\frac{p_{3}\hspace*{-0.35cm}/}{p_{12}^{2}-m_{Z_{H}}^{2}}P_{R}u(p_{1})
\\ \nonumber
&+&\frac{e^{2}x(1-3S_{W}^{2})(14-17S_{W}^{2})}{216S_{W}^{2}C_{W}^{2}cos2\theta_{W}}
\bar{v}(p_{2})\frac{p_{4}\hspace*{-0.35cm}/}{p_{12}^{2}-m_{Z_{H}}^{2}}P_{L}u(p_{1})
\\ \nonumber
&+&\frac{-e^{2}x(4-S_{W}^{2})(1-3S_{W}^{2})}{216S_{W}^{2}C_{W}^{2}cos2\theta_{W}}
\bar{v}(p_{2})\frac{p_{4}\hspace*{-0.35cm}/}{p_{12}^{2}-m_{Z_{H}}^{2}}P_{R}u(p_{1}).
\\\nonumber
\end{eqnarray}

\vspace*{-0.5cm}For the s-channel partonic processes
$q\bar{q}\rightarrow Z(Z_{H})\rightarrow \phi^{0}h$ $(q=d$, $s$ and
$b$), the invariant amplitude can be written\vspace{-0.8cm}

\begin{eqnarray}
\nonumber
M_{22}(q)&=&(\frac{e^{2}x}{24S_{W}^{2}C_{W}^{2}}-\frac{e^{2}x}{18C_{W}^{2}})
\bar{v}(p_{2})\frac{p_{12}\hspace*{-0.5cm}/}{p_{12}^{2}-m_{Z}^{2}}P_{L}u(p_{1})
\\ \nonumber
&+&\frac{e^{2}x}{36C_{W}^{2}}\bar{v}(p_{2})\frac{p_{12}\hspace*{-0.5cm}/}{p_{12}^{2}-m_{Z}^{2}}P_{R}u(p_{1})
\\ \nonumber
&+&\frac{e^{2}x(14-17S_{W}^{2})}{108C_{W}^{2}cos2\theta_{W}}
\bar{v}(p_{2})\frac{p_{3}\hspace*{-0.35cm}/}{p_{12}^{2}-m_{Z_{H}}^{2}}P_{L}u(p_{1})
\\ \nonumber
&+&\frac{e^{2}x(4-S_{W}^{2})(3-5S_{W}^{2})}{108S_{W}^{2}C_{W}^{2}cos2\theta_{W}}
\bar{v}(p_{2})\frac{p_{3}\hspace*{-0.35cm}/}{p_{12}^{2}-m_{Z_{H}}^{2}}P_{R}u(p_{1})
\\ \nonumber
&+&\frac{-e^{2}x(4-3S_{W}^{2})}{108C_{W}^{2}cos2\theta_{W}}
\bar{v}(p_{2})\frac{p_{4}\hspace*{-0.35cm}/}{p_{12}^{2}-m_{Z_{H}}^{2}}P_{L}u(p_{1})
\\ \nonumber
&+&\frac{-e^{2}x(4-S_{W}^{2})(3-5S_{W}^{2})}{108S_{W}^{2}C_{W}^{2}cos2\theta_{W}}
\bar{v}(p_{2})\frac{p_{4}\hspace*{-0.35cm}/}{p_{12}^{2}-m_{Z_{H}}^{2}}P_{R}u(p_{1}).
\\ \nonumber
\end{eqnarray}

\vspace*{-0.5cm}For the t-channel partonic process
$b\bar{b}\rightarrow \phi^{0}h$ as shown in $Fig.$7b, the invariant
amplitude can be written\vspace{-0.5cm}
\begin{eqnarray}
\nonumber M_{23}(q)&=&\frac{m_{b}^{2}}{\sqrt{2}v
f}\bar{v}(p_{2})\frac{p_{13}\hspace*{-0.5cm}/\hspace*{0.3cm}+m_{b}}{p_{13}^{2}-m_{b}^{2}}\gamma_{5}u(p_{1}).
\nonumber
\end{eqnarray}
Where $p_{13}=p_{1}-p_{3}$. Considering the couplings of the neutral
Higgs boson $\phi^{0}$ to the $SM$ fermions are proportional to the
factor of $m_{q}/f$ and the smallness masses of the quark $q=u, c,
d,$ and $s$, we have neglected their contributions to production
cross section of the neutral Higgs boson pair $\phi^{0}h$ via the
t-channel process in our calculations. When we calculate the loop
diagrams $Figs.$7(c)-7(d), and $Fig.$7e, we will use the same method
with $Figs.$4(a) and 4(b).

To see contributions of the different partonic processes to the
total hadronic cross section, we plot the total and partial hadronic
cross sections for different partonic processes as functions of the
parameter $f$ for $m_{\phi^{0}}=m_{h}=120GeV$ and $M=150GeV$ in
$Fig.$8. We see that the production cross sections of the neutral
Higgs bosons $\phi^{0}h$ mainly come from the contributions of the
light quarks ($u, d, c, s$) through the s-channel $Z$ exchange and
$Z_{H}$ exchange. Our numerical results show that, the contributions
coming from the partonic processes $gg\rightarrow \phi^{0}h$
[including $Figs.$7(c)-7(e)] to total production cross section are
at the orders of $10^{-5}fb- 10^{-1}fb$, which are much smaller than
those of the tree-level processes. This is because the $Yukawa$
couplings depend sensitively on the free parameters $M$ and $f$. The
parameter $M$ is very smaller than the scale parameter $f$. So,
although the gluon fusion get an enhancement due to large parton
distribution functions, the contribution of the gluon fusion process
is suppressed by the order of $(M/f)^{4}$ [21]. Thus, in $Fig.$8, we
did not show the line corresponding to the value of the production
cross section contributed by the $gg$ fusion. The value of the
production cross section of the neutral Higgs bosons $\phi^{0}h$ is
insensitive to the  mixing parameter $M$. For
$m_{\phi^{0}}=m_{h}=120GeV$ and $500GeV \leq f \leq 150GeV $, its
value is in the range of $34fb- 306fb$, the partial value of the
total production cross section coming from light quarks
contributions is in the range of $31fb- 281fb$. If we assume the
integrated luminosity $\pounds_{int}=100fb^{-1}$ for the $LHC$ with
the $c.m.$ energy $\sqrt{s}=14TeV$, then there will be
$3.4\times10^{3}- 3.1\times10^{4}$ $\phi^{0}h$ events generated at
the $LHC$.

\begin{figure}[htb]
\vspace{-0.9cm}
\begin{center}
\epsfig{file=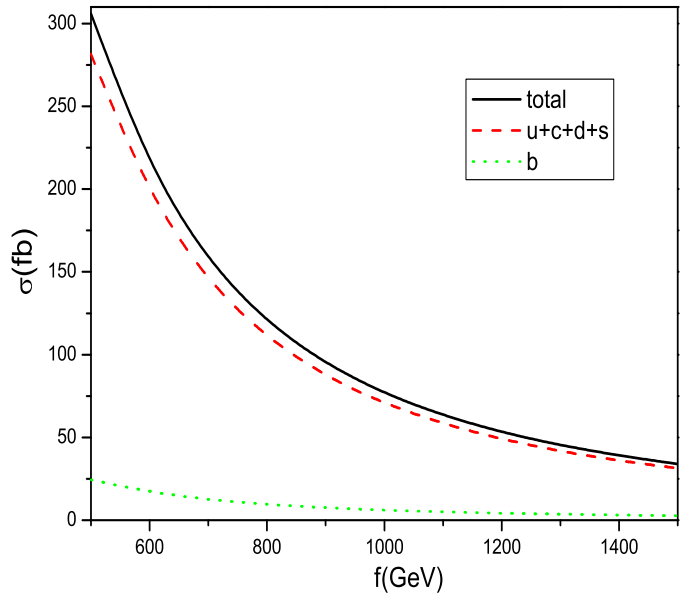,width=300pt,height=280pt}
\vspace{-1.1cm}\hspace{-0.7cm} \caption{The total and partial
hadronic cross sections for different partonic processes
\hspace*{2cm}as function of the parameter $f$  for
$m_{\phi^{0}}=m_{h}=120GeV$ and $M=150GeV$.} \label{ee}
\end{center}
\vspace{-0.6cm}
\end{figure}

Similar to those of the discussions for neutral Higgs boson pair
$\phi^{0}\phi^{0}$ production, we plot $\sigma$ as a function of
free parameter $f$ for $m_{h}=120GeV$, $M=150GeV$ and three values
of $m_{\phi^{0}}$ in $Fig.$9. One can see from $Fig.$9 that the
total cross section $\sigma$ is sensitive to mass parameter
$m_{\phi^{0}}$. For $f=500GeV$ and $120GeV\leq m_{\phi^{0}}
\leq180GeV$, its value is in the range of $101fb- 306fb$.

From the above discussions, we can see that the decay features of
$\phi^{0}$ are similar to those of the $SM$-like neutral Higgs boson
$h$, as far as decays into $b\bar{b}$ and $\gamma\gamma$ are
concerned. Therefore, when we analyze the signatures of the neutral
Higgs boson pairs from the $LRTH$ model at the colliders, we will
take the $\phi^{0}\phi^{0}$ pair, for example.

\begin{figure}[htb]
\vspace{-1cm}
\begin{center}
\epsfig{file=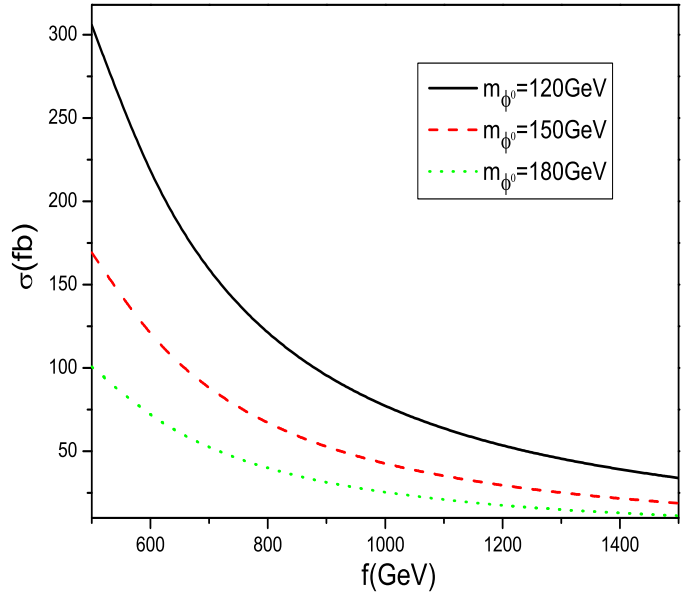,width=300pt,height=280pt}
\vspace{-1cm}\hspace{-0.5cm} \caption{The total production cross
section as a function of free parameter $f$ for
$m_{h}=\hspace*{1.7cm}120GeV$, $M=150GeV$ and three values of
$m_{\phi^{0}}$.} \label{ee}
\end{center}
\vspace{-0.6cm}
\end{figure}

In most of the parameter space of the $LRTH$ model, the main decay
modes of $\phi^{0}$ are $gg$ and $b\bar{b}$. However, the final
states $gggg$ and $b\bar{b}b\bar{b}$ induced by pair production of
the neutral Higgs boson $\phi^{0}$ at the $LHC$ have large $QCD$
backgrounds and thus are insignificant for $\phi^{0}$ discovery. If
we assume that one of the neutral Higgs boson $\phi^{0}$ decays to
$b\bar{b}$ and the other decays to $\gamma\gamma$, then pair
production of the neutral Higgs boson $\phi^{0}$ at the $LHC$ can
give rise to the $b\bar{b}\gamma\gamma$ final state, and the
production rate of the $b\bar{b}\gamma\gamma$ final state can be
easily estimated using the formula $\sigma_{s}=\sigma \times
Br(\phi^{0}\rightarrow b\bar{b})\times Br(\phi^{0}\rightarrow
\gamma\gamma)$. If we assume the integrated luminosity
$\pounds_{int}=100fb^{-1}$ for the $LHC$ with the $c.m.$ energy
$\sqrt{s}=14TeV$, then there will be several hundreds of
$b\bar{b}\gamma\gamma$ events to be generated at the $LHC$.
Furthermore, the narrow $\gamma\gamma$ peak can be reconstructed to
distinguish the signal from the backgrounds. Detailed analysis of
the signals and the relevant backgrounds about this kind of the
final state has been given in Ref.[36].

\section*{V. Conclusions}

\hspace{0.5cm}The twin Higgs mechanism provides an alternative
method to solve the little hierarchy problem. The $LRTH$ model is a
concrete realization of the twin Higgs mechanism. In this paper, we
discuss the possible decay modes of the neutral Higgs boson
$\phi^{0}$ predicted by the $LRTH$ model and consider its pair
production at the $ILC$ and $LHC$ via suitable mechanisms.

At the $ILC$, we study production of the neutral Higgs boson pair
$\phi^{0}h$ via the processes $e^{+}e^{-}\rightarrow Z(Z_{H})
\rightarrow \phi^{0}h$. Our numerical results show that, for
 $m_{\phi^{0}}=m_{h}=120GeV$ and $500GeV \leq f \leq 1500GeV $, the
total production cross section of neutral Higgs boson pair
$\phi^{0}h$ at $ILC$ is in the range of $0.92fb - 20fb$. If we
assume the integrated luminosity $\pounds_{int}=500fb^{-1}$ for the
$ILC$ with the $c.m.$ energy $\sqrt{s}=500GeV$, there will be
$10^{2}- 10^{4}$ $\phi^{0}h$ events to be generated at the $ILC$. If
we assume that the neutral Higgs bosons $\phi^{0}$ and $h$ both
decay to $b\bar{b}$, then the process $e^{+}e^{-}\rightarrow
\phi^{0}h$ can give rise to the $b\bar{b}b\bar{b}$ final state.
There will be $9- 3.0\times10^{3}$ $b\bar{b}b\bar{b}$ events to be
generated at the $ILC$. Owing to the $b\bar{b}b\bar{b}$ events, we
might detect the possible signatures of the neutral Higgs boson
$\phi^{0}$ via the processes $e^{+}e^{-}\rightarrow Z(Z_{H})
\rightarrow \phi^{0}h$ in the future $ILC$ experiments.

At the $LHC$, we study production of the neutral Higgs boson pairs
$\phi^{0}\phi^{0}$ and $\phi^{0}h$. First, we study production of
the neutral Higgs boson pair $\phi^{0}\phi^{0}$ via the processes
$gg\rightarrow \phi^{0}\phi^{0}$ and $q\bar{q}\rightarrow
\phi^{0}\phi^{0}$. Our numerical results show that,  for $M=100GeV$,
$m_{\phi^{0}}=120GeV$ and $ 500GeV \leq f \leq 1500GeV $, the value
of the hadronic cross section $\sigma_{\phi^{0}\phi^{0}}$ is in the
range of $4fb-122fb$, which mainly comes from the contributions of
the top-quark loop. Then we study production of the neutral Higgs
boson pair $\phi^{0}h$ via the processes $q\bar{q}\rightarrow
\phi^{0}h (q=u,c,d,s,b)$ and $gg\rightarrow \phi^{0}h$. Our
numerical results show that, for $M=150GeV$,
$m_{\phi^{0}}=m_{h}=120GeV$ and $ 500GeV \leq f \leq 1500GeV $, the
value of $\sigma_{\phi^{0}h}$ is in the range of $34fb - 306fb$, of
which about $91\%$ of the contributions comes from light quarks $u,
d, c, s$. If we assume the integrated luminosity
$\pounds_{int}=100fb^{-1}$ for the $LHC$ with the $c.m.$ energy
$\sqrt{s}=14TeV$, then there will be $3.4\times10^{3}-
3.1\times10^{4}$ $\phi^{0}h$ events to be generated at the $LHC$. If
we assume that one of the neutral Higgs bosons $\phi^{0}$ and $h$
decays to $b\bar{b}$ and the other decays to $\gamma\gamma$, then
the processes $pp\rightarrow \phi^{0}\phi^{0}+X$ and $pp\rightarrow
\phi^{0}h+X$ all can give rise to the $b\bar{b}\gamma\gamma$ final
state. There will be several hundreds and up to thousands of
$b\bar{b}\gamma\gamma$ events to be generated at the $LHC$ with the
$c.m.$ energy $\sqrt{s}=14TeV$ and $\pounds_{int}=100fb^{-1}$.

\section*{Acknowledgments}
\hspace{5mm}This work was supported in part by the National Natural
Science Foundation of China under Grants No.10675057, Specialized
Research Fund for the Doctoral Program of Higher Education(SRFDP)
(No.200801650002), the Natural Science Foundation of the Liaoning
Scientific Committee(No.20082148), and the Foundation of Liaoning
Educational Committee(No.2007T086). 


\vspace{1.0cm}


\begin{thebibliography}{99}
\bibitem{y1}For example see: N. E. Adam {\em et al.}, \emph{arXiv:} \textbf{0803.1154}[hep-ph].

\bibitem{y2}U. Baur, T. Plehn, and D. L. Rainwater, {\em Phys. Rev. Lett.} {\bf 89},
            151801(2002).
\bibitem{y3}M. Moretti, S. Moretti, F. Piccinini, R. Pittau, A. D. Polosa, {\em JHEP} {\bf0502},
             024(2005); T. Binoth, S. Karg, N. Kauer, R. Ruckl, {\em Phys. Rev. D}{\bf 74},
             113008(2006).
\bibitem{y4}A. Pierce, J. Thaler, Lian-Tao Wang, {\em JHEP} {\bf0705},
070(2007); S. Kanemura, K. Tsumura, \emph{arXiv:}
\textbf{0810.0433}[hep-ph].

\bibitem{y5}E. W. N. Glover and J. J. van der Bij, {\em Nucl. Phys. B}{\bf309},
282(1988).
\bibitem{y6}U. Baur, T. Plehn, David L. Rainwater, {\em Phys. Rev. D}{\bf67},
033003(2003).
\bibitem{y7}S. Dawson, S. Dittmaier, M. Spira, {\em Phys. Rev. D}{\bf58}, 115012(1998).
\bibitem{y8}S. Dawson, C. Kao, Yili Wang, P. Williams,
{\em Phys. Rev. D}{\bf75}, 013007(2007).


\bibitem{y9}J. J. Liu, W. G. Ma, G. Li, R. Y. Zhang and H. -S. Hou, {\em Phys. Rev. D}{\bf 70},
             015001(2004); C. O. Dib, R. Rosenfeld and A. Zerwekh, {\em JHEP} {\bf
             0605}, 074(2006); L. Wang, W. Y. Wang, J. M. Yang, H. J. Zhang, {\em Phys. Rev. D}{\bf76},
             017702(2007).

\bibitem{y10}P. K. Das and B. Mukhopadhyaya, {\em hep-ph}/{\bf0303135}.
\bibitem{y11}M. Spira and J. D. Wells, {\em Nucl. Phys. B}{\bf
            523}, 3(1998).
\bibitem{y12}A. A. Barrientos Bendezu, Bernd A. Kniehl, {\em Phys. Rev. D}{\bf64},
035006(2001).
\bibitem{y13}T. Plehn, M. Spira and P. M. Zerwas, {\em Nucl. Phys. B}{\bf479},
46(1996); A. Djouadi, W. Kilian, M. Muhlleitner and P. M. Zerwas,
            {\em Eur. Phys. J. C}{\bf10}, 45(1999);
             A. Belyaev, Manuel Drees, Oscar J. P. Eboli, J. K.
            Mizukoshi, S. F. Novaes, {\em Phys. Rev. D}{\bf60}, 075008(1999); A. Belyaev, M. Drees and J. K.
            Mizukoshi, {\em Eur. Phys. J. C}{\bf17}, 337(2000); R. Lafaye, D. J. Miller, M. Muhlleitner and S. Moretti,
            {\em hep-ph}/{\bf0002238}; M. Moretti, S. Moretti, F. Piccinini, R. Pittau, {\em JHEP} {\bf0502}, 024(2005).

\bibitem{y14}H. de Sandes, R. Rosenfeld, {\em Phys. Lett. B}{\bf
            659}, 323(2008).
\bibitem{y15}R. Barbieri and A. Strumia, {\em Phys. Lett. B}{\bf462},
144(1999); A. Falkowski, S. Pokorski, M. Schmaltz, {\em Phys. Rev.
D}{\bf74}, 035003(2006); Z. Chacko, H. -S. Goh, R. Harnik, {\em
Phys. Rev. Lett.} {\bf96}, 231802(2006).

\bibitem{y16}Z. Chacko, H. -S. Goh and R. Harnik, {\em JHEP} {\bf0601}, 108(2006).

\bibitem{y17}H. -S. Goh and S. Su, {\em Phys. Rev. D}{\bf75}, 075010(2007).

\bibitem{y18}A. Abada, I. Hidalgo, {\em Phys. Rev. D}{\bf77},
             113013(2008).
\bibitem{y19}D. -W. Jung and J. Y. Lee,
\emph{arXiv:} \textbf{0710.2589}[hep-ph].

\bibitem{y20}E. M. Dolle, S. F. Su, {\em Phys. Rev. D}{\bf77}, 075013(2008).
\bibitem{y21}Y. B. Liu , H. M. Han , X. L. Wang, {\em Eur. Phys. J. C}{\bf53}, 615(2008).

\bibitem{y22}J. F. Gunion, H. E. Haber, G. L. Kane, and S. Dawson,
             "The Higgs Hunter's Guide", Addison-Wesley, Reading,
             MA(1990); L. Reina, {\em hep-ph}/{\bf0512377}.



\bibitem{y23}W. -M. Yao {\em et al.} [Particle Data Group], {\em J. Phys. G}{\bf33}, 1(2006) and
partial updat for the 2008 edition.

\bibitem{y24}P. W. Higgs, {\em Phys. Rev. Lett.} {\bf13}, 508(1964);
G. S. Guralnik, C. R. Hagen and T. W. B. Kibble, {\em Phys. Rev.
Lett.} {\bf13}, 585(1964); F. Englert, R. Brout, {\em Phys. Rev.
Lett.} {\bf13}, 321(1964).

\bibitem{y25}G. Weiglein {\em et al.} [ILC/LC Study Group], {\em Phys.
Rept.} {\bf426}, 47(2006); A. Arhrib, R. Benbrik, C. -H. Chen, Rui
Santos,  \emph{arXiv:} \textbf{0901.3380}[hep-ph].

\bibitem{y26}J. J. Lopez-Villarejo, J. A. M. Vermaseren, \emph{arXiv:} \textbf{0812.3750}[hep-ph].

\bibitem{y27}A. Djouadi, V. Driesen, C. Junger, {\em Phys. Rev. D}{\bf54}, 759(1996).

\bibitem{y28}A. Gutierrez-Rodriguez, M. A. Hernandez-Ruiz, O. A. Sampayo, {\em Phys. Rev. D}{\bf67}, 074018(2003).
\bibitem{y29}H. Grosse, Yi Liao, {\em Phys. Rev. D}{\bf64},
115007(2001).
\bibitem{y30}J. L. Feng, T. Moroi, {\em Phys. Rev. D}{\bf56}, 5962(1997).

\bibitem{y31}A. Djouadi, H. E. Haber, P. M. Zerwas, {\em Phys. Lett. B}{\bf375}, 203(1996).

\bibitem{y32}R. N. Hodgkinson, D. Lopez-Val, Joan Sola, {\em Phys. Lett. B}{\bf673},
47(2009); A. Arhrib, R. Benbrik, C. W. Chiang, {\em Phys. Rev.
D}{\bf77}, 115013(2008).


\bibitem{y33}http://www.linearcollider.org/newsline/pdfs/20061207$_{-}$LC$_{-}$Parameters$_{-}$Novfinal.pdf


\bibitem{y34}T. Hahn, M. Perez-Victoria, {\em Computl. Phys. Commun. }{\bf118}, 153(1999);
T. Hahn, {\em Nucl. Phys. Proc. Suppl.} {\bf135}, 333(2004).

\bibitem{y35}J. Pumplin {\em et al.} (CTEQ Collaboration), {\em JHEP} {\bf0602}, 032(2006).

\bibitem{y36}U. Baur, T. Plehn , David L. Rainwater, {\em Phys. Rev. D}{\bf69}, 053004(2004).




\end{thebibliography}
\end{document}